\documentclass[twocolappendix,twocolumn]{aastex631}
\usepackage[T1]{fontenc}
\usepackage{ae,aecompl}
\usepackage{bethmacros}
\usepackage{graphicx}
\usepackage{amssymb}
\usepackage{amsmath}
\usepackage{placeins}
\usepackage{times}

\def\spose#1{\hbox to 0pt{#1\hss}}
\def\lta{\mathrel{\spose{\lower 3pt\hbox{$\mathchar"218$}}
     \raise 2.0pt\hbox{$\mathchar"13C$}}}
\def\gta{\mathrel{\spose{\lower 3pt\hbox{$\mathchar"218$}}
     \raise 2.0pt\hbox{$\mathchar"13E$}}}

\newcommand{\appropto}{\mathrel{\vcenter{
  \offinterlineskip\halign{\hfil$##$\cr
    \propto\cr\noalign{\kern2pt}\sim\cr\noalign{\kern-2pt}}}}}

\shorttitle{Can Cosmological Simulations Reproduce Observed $z\geq 10$ Galaxies}
\shortauthors{B.W. Keller}

\begin{document}
\title{Can Cosmological Simulations Produce the Spectroscopically Confirmed
Galaxies Seen at $z\geq 10$?}
\author[0000-0002-9642-7193]{B.W. Keller}
\affiliation{Department of Physics and Materials Science, University of Memphis,
\\ 3720 Alumni Avenue, Memphis, TN 38152, USA}
\email{bkeller1@memphis.edu}
\author[0000-0002-9581-0297]{F. Munshi}
\affiliation{Department of Physics and Astronomy, George Mason University \\ 4400 University Drive, MSN 3F3
Fairfax, VA 22030-4444, USA}
\author[0000-0002-6849-5375]{M. Trebitsch}
\affiliation{Kapteyn Astronomical Institute, University of Groningen, \\ P.O. Box
800, 9700 AV Groningen, The Netherlands}
\author[0000-0002-4353-0306]{M. Tremmel}
\affiliation{Physics Department, University College Cork, T12 K8AF Cork,
Ireland}
\begin{abstract} 
    Recent photometric detections of extreme $(z>10)$ redshift galaxies from the
    JWST have been shown to be in strong tension with existing simulation models
    for galaxy formation, and in the most acute case, in tension with $\Lambda
    CDM$ itself.  These results, however, all rest on the confirmation of these
    distances by spectroscopy.  Recently, the JADES survey has detected the most
    distant galaxies with spectroscopically confirmed redshifts, with four
    galaxies found with redshifts between $z=10.38$ and $z=13.2$.  In this
    paper, we compare simulation predictions from four large cosmological
    volumes and two zoom-in protoclusters with the JADES observations to determine
    whether these spectroscopically confirmed galaxy detections are in tension
    with existing models for galaxy formation, or with $\Lambda CDM$ more
    broadly.  We find that existing models for cosmological galaxy formation can
    generally reproduce the observations for JADES, in terms of galaxy stellar
    masses, star formation rates, and the number density of galaxies at $z>10$.
\end{abstract}


\section{Introduction}
The successful deployment of the JWST has already produced observations of the
highest-redshift galaxies detected to date.  The first sets of detections
reported \citep{Finkelstein2022,Labbe2022,Naidu2022a,Adams2023}
have found galaxies with $M_*>10^8\Msun$ at $z_{phot}>10$.  The most extreme of
these examples, with $M_*>10^{10}\Msun$ at $z_{phot}>10$ have already been shown
to be in strong tension with $\Lambda CDM$
\citep{Haslbauer2022,Boylan-Kolchin2022}.  These tensions have, however, been
clouded by the large uncertainty in fitting photometric redshifts at such
extreme distances \citep{Bouwens2022,Naidu2022b,Kaasinen2022}.
\citet{Naidu2022b} and \citet{Zavala2022} found that breaks in the Spectral
Energy Distribution (SED) produced by dust obscuration at $z\sim5$ can
masquerade as a Lyman break at $z\sim17$ for recent JWST observations (see also
\citep{Fujimoto2022} for a similar effect in ALMA observations). 

As \citet{Boylan-Kolchin2022} showed, if the comoving number density of galaxies
at $z=10$ with $M_*>10^{10}\Msun$ is as high as can be estimated from
\citet{Labbe2022}, it would be a significant challenge for $\Lambda CDM$:
analogous to Haldane's ``fossil rabbits in the Precambrian'' \citep{Harvey1996}.
\citet{Haslbauer2022} has shown that the observations of galaxies with high
stellar mass ($M_*>10^9$) at high redshift $z\gtrsim10$ from \citet{Adams2023},
\citet{Labbe2022}, \citet{Naidu2022a}, and \citet{Naidu2022b} are in extreme
tension with the simulation predictions of EAGLE \citep{Schaye2015}, TNG50, and
TNG100 \citep{Pillepich2018a, Nelson2019b}.  These authors warn however that
spectroscopic confirmation of redshifts is needed before final conclusions can
be reached: these tensions all rest on the estimations of stellar masses
provided by SED fitting and, critically, on the distance estimates themselves.
Without spectroscopic confirmation of the redshifts reported in these
observations, these potential tensions may be illusory: artifacts of
overestimated distances, and thus overestimated intrinsic luminosities.  Indeed,
as \citet{Behroozi2020} has shown, semi-empirical modelling of galaxy formation
in $\Lambda CDM$ predicts JWST-detectable galaxies with $M_*>10^7\Msun$ to at
least $z\sim13.5$.

Spectroscopic confirmation has now arrived with the discovery in the JADES
survey of 4 galaxies with $z_{spec}>10$ and $M_*\gtrsim10^8\Msun$
\citep{Curtis-Lake2022,Robertson2022}.  By observing 65 arcmin$^2$ of the
GOODS-S field with JWST NIRCam and NIRSpec, JADES has confirmed the four
earliest detected galaxies: JADES-GS-z10-0 at $z=10.38^{+0.07}_{-0.06}$,
JADES-GS-z11-0 at $z=11.58^{+0.05}_{-0.05}$, JADES-GS-z12-0 at
$z=12.63^{+0.24}_{-0.08}$, and JADES-GS-z13-0 at $z=13.2^{+0.04}_{-0.07}$.
\citet{Curtis-Lake2022,Robertson2022} have measured stellar masses and star
formation rates for these galaxies using the \textit{Prospector} SED-fitting
code \citep{Johnson2021}, finding them to be compact, star forming galaxies with
young stellar populations and relatively high star formation surface densities.

In this letter, we compare the observed JADES galaxies to predictions from an
array of large cosmological hydrodynamical simulations.  These simulations
reproduce observed galaxy population statistics (such as the observed galaxy
stellar mass function and fundamental plane of star formation) at low redshift.
With JADES, we are able to test whether those same models for galaxy formation
fail to reproduce these new observations of high redshift galaxy formation.

\section{Simulation Data}
\begin{table*}
    \centering
    \begin{tabular}{l | l l l l l l}
        \hline
        Simulation & Cosmology & Box Size (cMpc) & z range & $N_{snap}$ & $M_{DM} (\Msun)$ &
        $M_{baryon} (\Msun)$ \\
        \hline
        \hline
        EAGLE & Planck 2013 \citep{Planck2014} & 100 & 9.99 & 1 & $9.7\times10^6$ & $1.81\times10^6$\\
        Illustris & WMAP-9 \citep{Hinshaw2013} & 106.5 & 10.00-13.34 & 7 & $6.3\times10^6$ & $1.3\times10^6$ \\
        TNG100 & Planck 2015 \citep{Planck2016} & 110.7 & 10.00-11.98 & 3 & $7.5\times10^6$ &
        $1.4\times10^6$ \\
        Simba & Planck 2015 \citep{Planck2016} & 147.7 & 9.96-13.70 & 10 & $9.7\times10^7$ &
        $1.82\times10^7$ \\
        OBELISK & WMAP-7 \citep{Komatsu2011} & 142.0$^\dag$ & 10.07-13.77 & 13 &
        $1.2\times10^6$ & $1\times10^4$ \\
        RomulusC & Planck 2015 \citep{Planck2016} & 50$^\dag$ & 9.97-12.88 & 2 &
        $3.4\times10^5$ & $2.1\times10^5$ \\
    \end{tabular}
     \caption{Simulation data compared in this study.  Each cosmological volume
     has a box of side length $\sim100 \cMpc$.  We show the choice of
     cosmological parameters, box size, redshift range between $z\sim10-14$,
     number of snapshot outputs in that range, and resolution for DM and
     baryons.  The ($^\dag$) denotes that these are zoom-in simulations of an
     individual galaxy protocluster, drawn from a lower-resolution volume.
     OBELISK is an Eulerian simulation, and thus the baryonic resolution
     reported here is given as the typical mass of a star particle and the mass
     of a gas cell with $\Delta x=35\pc$ at a density of $n>10\hcc$.}
    \label{simulations}
\end{table*}
In order to probe the predictions of current galaxy formation models, we examine
simulation data from the EAGLE \citep{Crain2015,Schaye2015,McAlpine2016},
Illustris \citep{Vogelsberger2014a,Vogelsberger2014b,Genel2014,Nelson2015b},
TNG100
\citep{Pillepich2018b,Springel2018,Nelson2018,Naiman2018,Marinacci2018,Nelson2019a},
and Simba \citep{Dave2019} cosmological volumes.  Each of these simulations has
a volume of $\sim10^6\cMpc^3$, with baryonic mass resolution of
$10^6\Msun-10^7\Msun$, which allows them to (marginally) resolve the formation
of $M_*=10^8\Msun-10^9\Msun$ galaxies (an $M_*=5\times10^8\Msun$ galaxy in EAGLE,
Illustris, TNG100, and Simba will contain 276, 384, 357, and 27 star particles
respectively).  Each of these projects includes models
for gas cooling, star formation, and feedback from supernovae (SNe) and active
galactic nuclei (AGN) that are tuned to reproduce the $z\sim0$ stellar mass
function (among other low-redshift population statistics).  In
Table~\ref{simulations} we list the cosmological parameters used for each
simulation, as well as the size of the simulation volume, the range and number
of snapshots with redshift between $z\sim10$ and $z\sim14$, and the mass
resolution for dark matter (DM) and baryonic particles.  We rely on the public
releases of the halo catalogs from each simulation to determine galaxy stellar
masses and star formation rates in these simulation data sets.

We have also included data from the cosmological zoom-in simulations OBELISK
\citep{Trebitsch2021} and RomulusC \citep{Tremmel2019}.  OBELISK is a zoom-in of
the most massive halo at $z=2$ in the {\sc Horizon-AGN} volume
\citep{Dubois2014}.  This region is selected to include all DM
particles within $4R_{vir}$ of the halo center at $z=2$.  At $z=0$, this cluster
has a halo mass of $M_{halo}\sim 6.6\times10^{14}\Msun$.  RomulusC is a zoom-in
simulation of a smaller galaxy cluster, with halo mass
$M_{halo}=1.5\times10^{14}\Msun$ at $z=0$.  It is drawn from a $50^3 \cMpc^3$
DM-only volume, and applies the same modelling approach for cooling, star
formation, supermassive black hole (SMBH) evolution, and feedback as the Romulus volume
\citep{Tremmel2017}.  Unlike the other data sets we examine here, these zooms do
not include a full resolution sample of a large volume; instead they offer an
higher-resolution picture of early collapsing overdensities.

\begin{figure}
    \includegraphics[width=0.5\textwidth]{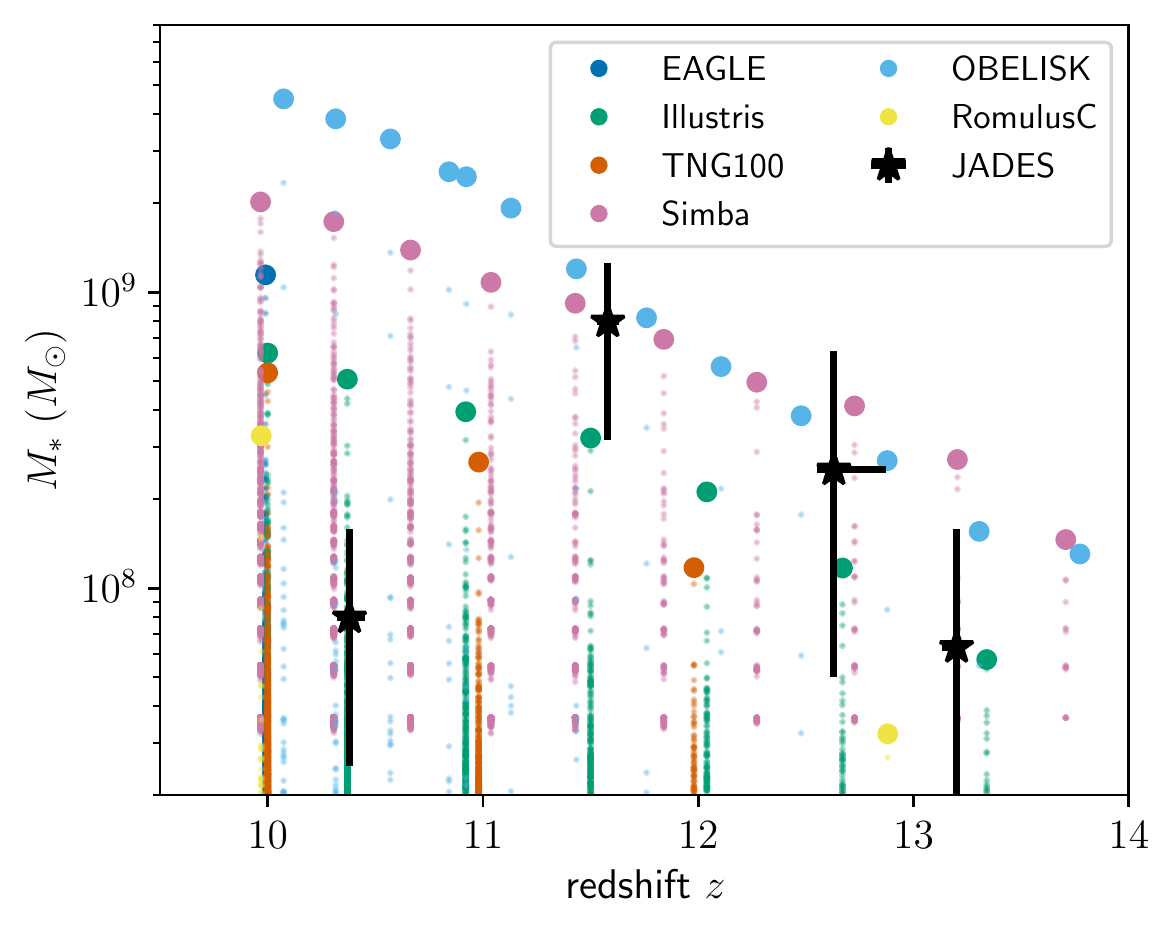}
    \caption{Stellar mass of galaxies from various simulation volumes as a
    function of redshift.  Black stars with error bars show the
    \citet{Robertson2022} JADES observations, while colored points show
    individual galaxies from different simulated volumes.  Large colored points show the
    most massive galaxy at each redshift for a given simulation.}
    \label{Mstar_z}
\end{figure}

\begin{figure}
    \includegraphics[width=0.5\textwidth]{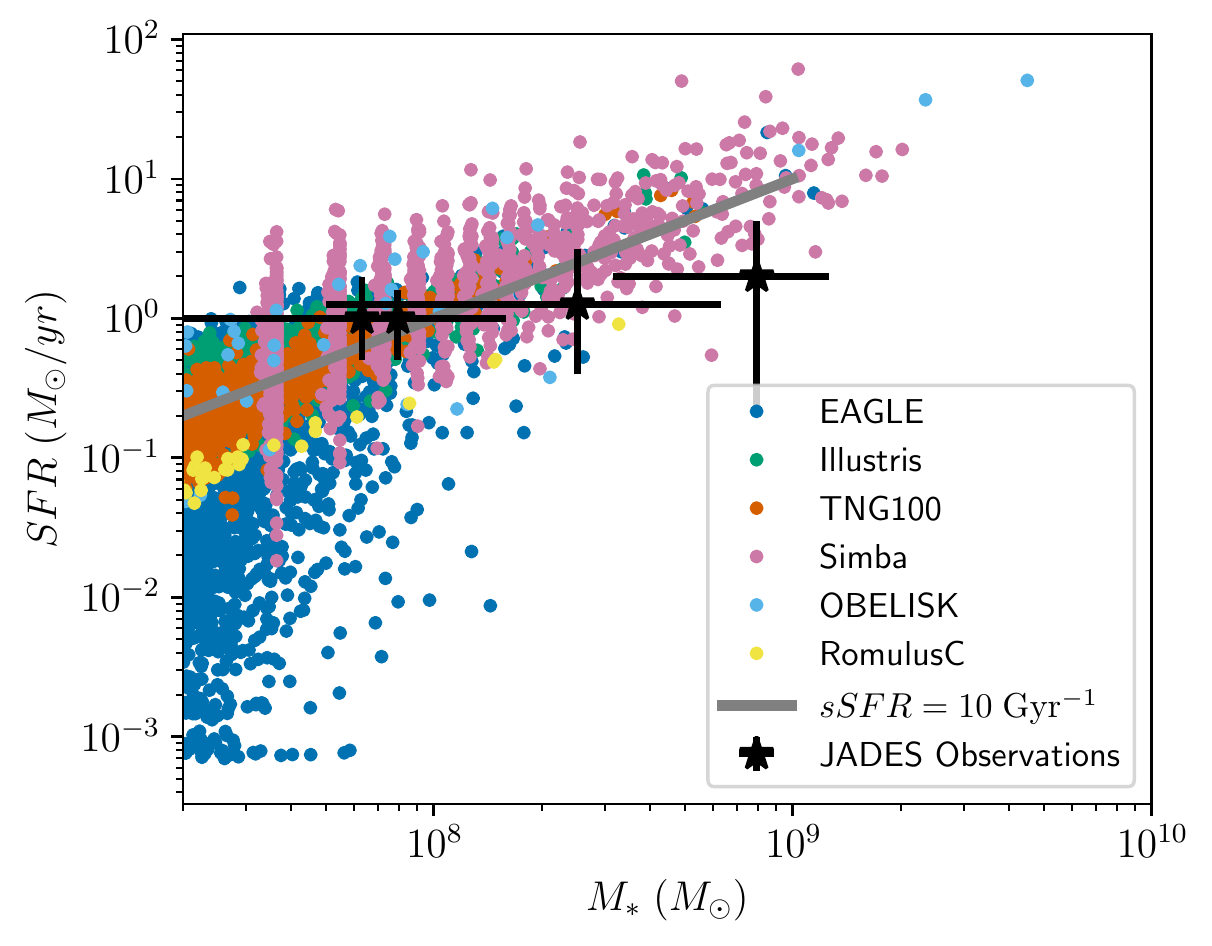}
    \caption{Star formation rate versus stellar mass for simulated galaxies at
    $z\sim10$ and observed galaxies at $z\geq10$.  Simulation data are shown as
    colored points, while the JADES observations are shown as black stars.  A
    constant sSFR of $10\Gyr^{-1}$ is shown in the grey line.}
    \label{SFR_Mstar}
\end{figure}

\begin{figure*}
    \includegraphics[width=\textwidth]{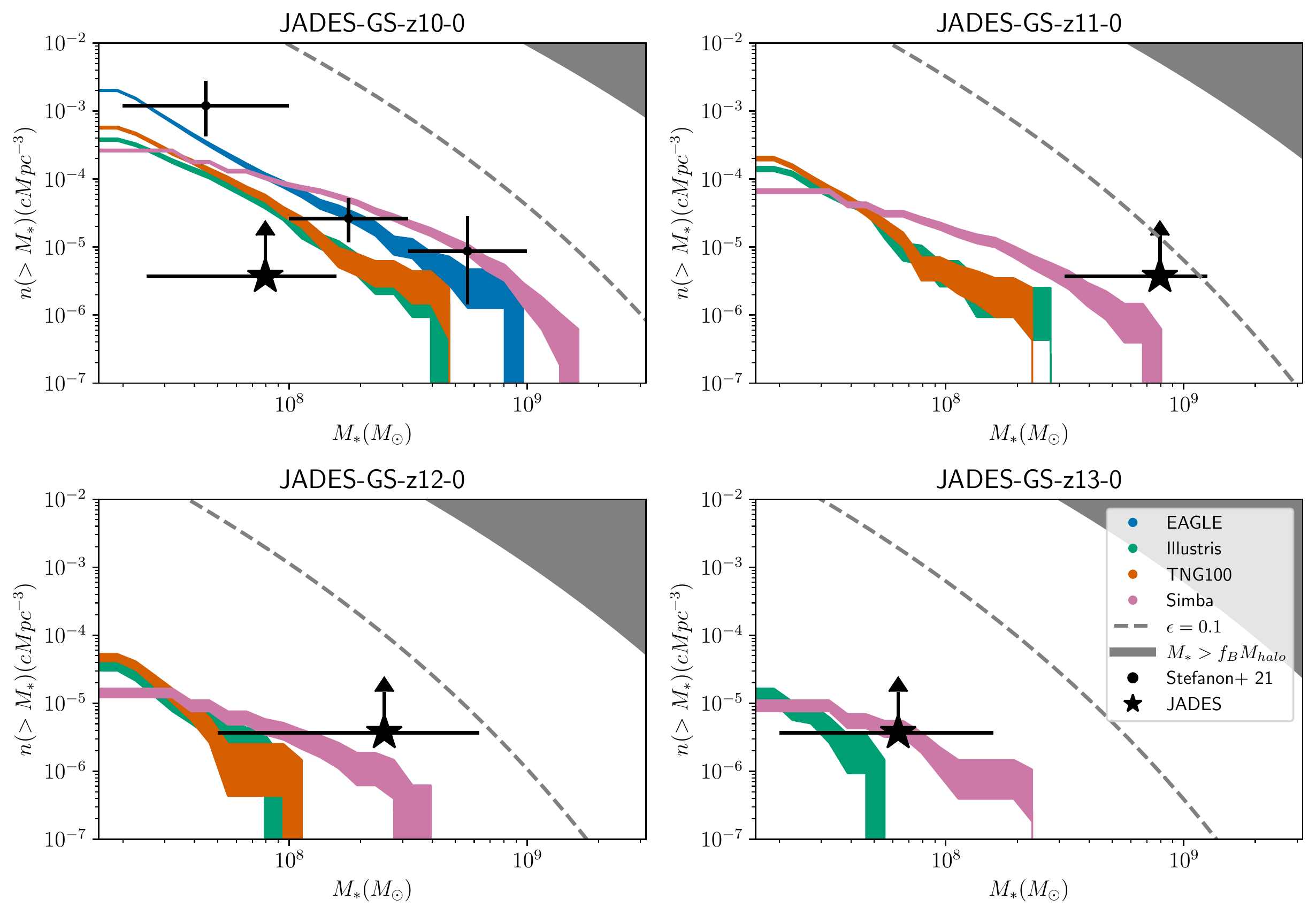}
    \caption{Cumulative number density of galaxies above a given stellar mass
    for simulation snapshots (colored areas) at the nearest redshift to the
    JADES observations.  Snapshots are chosen such that no more than $\Delta z
    =0.5$ separates the simulation snapshot redshift from the observed redshift.
    Black stars show the detections from \citet{Robertson2022}. Filled colored
    areas show the uncertainty from Poisson sampling of each mass bin.  The grey
    dashed curve shows the expected number density for an integrated star
    formation efficiency $\epsilon$ of 10 per cent, and the shaded grey are
    shows the excluded region where more stellar mass is produced in halos than
    their available baryon budget.  Black circles at $z\sim10$ show the
    estimated number density from observations by \citet{Stefanon2021}.  We
    exclude from this figure the data from zoom-in simulations.}
    \label{num_den}
\end{figure*}

\section{Results}
We begin by simply showing the distribution of galaxy stellar masses in each
simulation volume as a function of redshift, shown in Figure~\ref{Mstar_z}.  Not
all of the simulations we examine here have well-sampled snapshots above $z=10$
(EAGLE in particular only includes one snapshot between $z\sim10$ and
$z\sim14$).  However, as can be seen, all of the simulations produce a large
number of galaxies with $M_*>10^8\Msun$ at $z\sim10$.  At all redshifts, Simba
and OBELISK produce the most massive galaxies of the simulations we examine, in
part due to the larger volume they simulate/are drawn from ($\sim2.4$ times the
comoving volume of TNG100, the next largest volume), as well as differences in
the choice of subgrid physics models.  As we move to higher redshift, only Simba
and OBELISK produce even a single galaxy above the best-fit stellar masses of
JADES-GS-z11-0, JADES-GS-z12-0, and JADES-GS-z13-0.  Illustris and TNG100 are
unable to produce galaxies at $z\sim11.5-12$ which reach even the lower estimate
for the stellar mass of JADES-GS-z11-0.

As \citet{Robertson2022} has also provided estimates for the star formation
rates (SFR) in the 4 JADES observations, we show in Figure~\ref{SFR_Mstar} the
SFR-$M_*$ plane for simulated galaxies at $z\sim10$.  All simulation volumes
produce galaxies at $z\sim10$ with $sSFR=10\Gyr^{-1}$, except for RomulusC,
which has an $sSFR$ roughly half this value. An $sSFR$ of $10\Gyr^{-1}$ is
approximately 4 times higher than the star forming main sequence at $z=2$
\citep{Rodighiero2011},  consistent with the observed trend of increasing sSFR
towards higher $z\sim6$ redshift \citep{Santini2017}.  The banding
seen in the lower-mass Simba galaxies is simply a function of resolution, as
these galaxies contain only a handful of star particles.  Interestingly, there
appears to be no noticeable trend in the SFR as a function of stellar mass for
the JADES observations. However, drawing strong conclusions as to the nature of
the SFR$-M_*$ relation at $z\geq10$ is ill-advised given the small number of
observations, their relatively large uncertainties, and the intrinsic scatter in
the SFR-$M_*$ plane that the simulations predict.  Beyond these, the NIRSpec
observations of the JADES galaxies are selected from photometric observations,
which are subject to a continuum flux (and therefore SFR) limit, introducing a
potential observational bias.  Overall, the SFRs measured from the simulations
match the JADES observations reasonably well, though with perhaps higher SFRs
for simulated galaxies at masses above $M_*\sim5\times10^8\Msun$.  Further
observations will help reveal if the flat $SFR-M_*$ curve is a real feature of
$z\geq10$ galaxy formation, or simply an artifact of the small sample size of
the spectroscopically confirmed JADES galaxies and/or uncertainties in the
estimation of $M_*$ or $SFR$.

In order to make a more quantitative estimate of potential tension between
simulated cosmological volumes and the \citet{Robertson2022} observations, we
now look to the comoving number density of galaxies at each of the redshifts
probed by JADES. The deep observing area of JADES, using the JWST NIRCam, (9.7
square arcmin), yields a volume of $V\sim(9\times9 
\times494)\cMpc^3\sim4\times10^4\cMpc^3$ bounded by the comoving distance of
$494\Mpc$ between the JADES-GS-z10-0 at $z=10.38$ and JADES-GS-z13-0 at
$z=13.2$.  These candidates were, however, originally selected photometrically
from a wider field of 65 arcmin$^2$, which yields a volume of
$V\sim2.7\times10^5\cMpc^3$.  We can thus estimate the number density of the
JADES galaxy observations as $n\sim 3.7\times10^{-6}\cMpc^{-3}$.  In
Figure~\ref{num_den}, we show the comoving number density of galaxies above a
given stellar mass for simulation snapshots nearest in redshift to each of the
JADES observations (if the nearest snapshot is separated by more than 0.5 in
redshift, we omit it from the plot).  Each of the JADES
observations above $z>10.38$ implies a slightly higher number density than
what is produced by the simulations.  We also show the estimated number density
assuming a constant stellar baryon conversion efficiency ($M_*=\epsilon f_B
M_{halo}$) of $\epsilon=0.1$, and the maximum expected number density (with
$M_*=f_B M_{halo}$), following \citet{Boylan-Kolchin2022} using a
\citet{Sheth1999} halo mass function.  Each of the JADES galaxies lies outside
of the excluded region of $M_*=f_B M_{halo}$, but the most extreme two cases
(JADES-GS-z11-0 and JADES-GS-z12-0) imply integrated baryon conversion
efficiencies near 10 per cent ($M_*/(f_B M_{halo})\sim0.1$).  For $z\sim10$, we
also show the independent measurements for the galaxy stellar mass function
measured by the GREATS ALMA survey by \citet{Stefanon2021}.  The highest
predicted number density for massive halos at these redshifts is from Simba, and
only JADES-GS-z11-0 and JADES-GS-z12-0 imply number densities higher than in
Simba for galaxies at their stellar mass.  Even for the most extreme case,
JADES-GS-z11-0, the probability of finding a galaxy with stellar mass above
$M_*=10^{8.9}\Msun$ at $z\sim11$, in a random volume of $2.7\times 10^5\cMpc^3$
in Simba, is $8\%$.  If we instead take the lower estimate for the stellar masses
from JADES-GS-z11-0 ($M_*=10^{8.5}\Msun$), this probability rises to $92\%$.
There does not appear to be any significant tension in the density of galaxies
with $M_*\sim10^8\Msun$ at $z>10$ inferred from JADES and the number density
produced by at least one existing cosmological simulation (Simba).

\section{Discussion}
%Summarize results
We have compared the recent detection of spectroscopically confirmed $z>10$
galaxies from the JADES survey to the EAGLE, Illustris, TNG100, and Simba
surveys.  We show that all four simulation suites produce galaxies with
$M_*\sim10^{8.5}\Msun$ at $z\sim10$, and that Simba in particular produces at
least one galaxy with mass comparable to the JADES observations at each
observed redshift.  The JADES galaxies show a relatively flat $SFR-M_*$ trend, in
contrast to the constant sSFR of $\sim5-10\Gyr^{-1}$ for simulated galaxies at
$z\sim10$.  Comparing the estimated JADES galaxy stellar mass functions to the
simulation predictions shows that these simulations are in reasonable agreement
with the number density of galaxies at $z\sim10$ observed by JADES and the
earlier ALMA measurements from \citet{Stefanon2021} (though Illustris and TNG100
appear to be slightly under-predicting the formation of galaxies at $z\sim10$).
At $z\sim11-12$, JADES implies a slightly higher number density of galaxies at
$M>10^8\Msun$ than are predicted by any of the simulation volumes we examine
here.  We find, however, that given the small number of objects confirmed in
JADES (in a volume of $\sim2.7\times10^5\cMpc^3$), none of the JADES observations
is in greater than $2\sigma$ tension for the best-fit stellar mass, relative to the
Simba simulation.  JADES-GS-z11-0 implies a slightly higher density at $z=11$
for galaxies with $M_*\sim10^9\Msun$, but the difference between Simba and
density inferred from JADES is only slight. A randomly chosen volume of
$2.7\times10^5\cMpc^3$ from Simba at $z=11.4$ will contain a galaxy with stellar
mass greater than that of JADES-GS-z11-0 $8\%$ of the time.  At the lower
uncertainty for the JADES-GS-z11-0 stellar mass, this probability grows to
$92\%$.  

As future observations better constrain the stellar mass function at $z\sim11$,
this tension may strengthen or disappear.  If objects as massive as those seen
by \citep{Labbe2022,Adams2023,Naidu2022a} are confirmed to be common, with
spectroscopic redshifts of $z>10$, this would imply a serious problem with
existing $\Lambda CDM$ and galaxy formation theory
\citep{Boylan-Kolchin2022,Haslbauer2022,Lovell2023}.  In order to be in tension
with all models for galaxy formation in $\Lambda CDM$ (by implying galaxy star
formation efficiencies $>100\%$), the number density of $M_*=10^9\Msun$ galaxies
at $z\sim11$ would need to be more than 3 orders of magnitude higher than what
is implied by the JADES-GS-z11-0 detection.  This tension may be resolved by
future refinements in the SED-fit stellar mass estimates lowering the stellar
mass measured for JADES-GS-z11-0, or simply by JADES-GS-z11-0 residing in a
moderately rare ($P\sim8\%$) overdensity. For a mean cosmological mass density
of $\rho_m \sim 4\times10^{10}\Msun\cMpc^{-3}$, a comoving volume of
$4\times10^4\cMpc^3$ contains a total mass of $\sim 10^{15}\Msun$.  If the
entire volume covered by the deepest JADES observations is a rare overdensity
that will eventually collapse to form a $z\sim0$ cluster, it will be on the
order of the most massive clusters detected \citep{Lovell2023}.  It is unlikely
that the entire JADES volume is probing a single large overdensity (given the
fact that it is a pencil-beam volume with a comoving depth of $494\cMpc$ over an
area on the sky of only $(9\times9)\cMpc^2$), but the possibility that one or
more of the JADES galaxies above $z>10$ resides within one or more protoclusters
such as those simulated in OBELISK is still plausible.

% Why is Simba higher?
One question that arises from this data is why the Simba simulations predict
more $M_*>10^8\Msun$ galaxies at $z>10$ than Illustris and TNG100.  While the
Simba volume is somewhat larger than Illustris and TNG100 (Simba's volume is
$\sim2.4$ times larger than TNG100), it also produces a higher density of
galaxies with $M_*>10^8\Msun$ at all redshifts we have examined here.  Each of
these simulations applies a different set of models for star formation and
feedback by SNe and AGN. In Simba, SMBHs are seeded in galaxies with
$M_*>10^{9.5}\Msun$, while in TNG100 SMBHs are seeded in halos with
$M_{halo}>7.4\times10^{10}\Msun$.  This means that none of the galaxies we show
in Figure~\ref{Mstar_z} or in Figure~\ref{num_den} contain SMBHs, so the
differences we find must be a function of the different star formation and
feedback recipes used in TNG100 and Simba.  Simba applies a tuned reduction in
the SN-driven outflow mass loadings $\eta$ above $z>3$ to better match
observations of the galaxy stellar mass function at $z>6$, lowering the mass
loadings by $(a/0.25)^2$ \citep{Dave2019}.  It appears that lower mass loadings
at high redshift are not only important to matching observations at $z\sim6$,
but for $z>10$ as well.  Understanding how SNe regulate galaxy formation and
drive outflows in these early epochs will be an important avenue of future
simulation study.

%Compare to high-z (reionization) simulations
While the results we have shown here show that there is no strong tension
between at least one existing large-volume cosmological simulation and the
spectroscopically confirmed galaxy detections from JADES, it is important to
note that the cosmological volumes we have examined here are all relatively low
resolution: even a $10^9\Msun$ galaxy only contains $\sim50$ star particles in
Simba. These simulations are also tuned to reproduce $z\sim0$ galaxy properties.  Star
formation and feedback at low redshift are still relatively poorly understood
processes \citep{Naab2017}, a problem that becomes much more severe at epochs 
as early as $z>10$ \citep{Visbal2020}.

An obvious direction for future studies is to search for galaxies of similar
mass in higher resolution, high redshift simulation volumes.  In particular,
simulations designed to study reionization such as Renaissance
\citep{O'Shea2015}, OBELISK \citep{Trebitsch2021}, and SPHINX
\citep{Rosdahl2018}, all achieve resolutions significantly higher than any of
the volumes we have studied here. Many also feature more physically motivated
models for stellar feedback, which are possible at these higher resolutions.
Even at the same resolution as these $(\sim100\Mpc)^3$ volumes, rare regions can
be probed by zooming in on overdensities from much larger volumes, a strategy
applied by FLARES \citep{Lovell2021}.  FLARES produces a stellar mass function
at $z\sim10$ consistent with \citet{Stefanon2021}, which we show here to be
consistent with JADES.  Zoom-in simulations of protoclusters such as OBELISK
will be particularly fruitful, as the earliest bright galaxies to form are
likely to be found in these environments.  As we show in Figure~\ref{Mstar_z},
OBELISK contains a number of simulated galaxies with comparable stellar masses
to each of the JADES detections at every redshift of the JADES sample, resolved
with $>1000$ star particles.  Zoom-in simulations such as these will be a
powerful tool for understanding the earliest phase of galaxy formation at
$z>10$.

\section{Conclusions}
We have compared the recent observations of the highest redshift galaxies with
spectroscopically confirmed distances from the JADES survey
\citep{Robertson2022} to simulated galaxies from EAGLE, Illustris, TNG100, and
Simba volumes and the OBELISK and RomulusC zoom-ins.  In general, we find that each of these
simulations produces galaxies with comparable stellar masses to the JADES
galaxies by $z\sim10$.  The most massive JADES galaxies have somewhat lower SFRs
than simulated galaxies at $z\sim10$, but lie within the scatter of the
simulations.  The galaxy number density implied by the JADES galaxies at
$z\sim10$ is consistent with both the simulations and past observations.  At
higher redshift, only Simba and OBELISK produce galaxies as massive as are found
in JADES.  The number density of galaxies inferred from JADES is slightly larger
than what is predicted by Simba at $z=11$ and $z=12$, but at a low level of
significance.  Overall, there appears to be no strong tension between models for
galaxy formation in cosmological hydrodynamic simulations and the most distant
spectroscopically confirmed galaxies.

\section*{Acknowledgements}
BWK would like to thank Tessa Klettl for her incredible support and help editing
this manuscript.  M. Trebitsch acknowledges support from the NWO grant 0.16.VIDI.189.162
(``ODIN'').  We also would like to thank James Wadsley for useful discussions,
and for Jordan Van Nest in helping with the RomulusC simulation data. We
especially would like to thank the teams behind EAGLE, Illustris, TNG, and Simba
for providing public access to their simulation datasets.  This study would have
been impossible to complete in a timely fashion without the community spirit
behind these public releases.

We acknowledge the Virgo Consortium for making their simulation data available.
The EAGLE simulations were performed using the DiRAC-2 facility at Durham,
managed by the ICC, and the PRACE facility Curie based in France at TGCC, CEA,
Bruy\'eresle-Ch\^atel.

The IllustrisTNG simulations were undertaken with compute time awarded by the
Gauss Centre for Supercomputing (GCS) under GCS Large-Scale Projects GCS-ILLU
and GCS-DWAR on the GCS share of the supercomputer Hazel Hen at the High
Performance Computing Center Stuttgart (HLRS), as well as on the machines of the
Max Planck Computing and Data Facility (MPCDF) in Garching, Germany.

\bibliographystyle{aasjournal}
\bibliography{references}

\begin{thebibliography}{}
\expandafter\ifx\csname natexlab\endcsname\relax\def\natexlab#1{#1}\fi
\providecommand{\url}[1]{\href{#1}{#1}}
\providecommand{\dodoi}[1]{doi:~\href{http://doi.org/#1}{\nolinkurl{#1}}}
\providecommand{\doeprint}[1]{\href{http://ascl.net/#1}{\nolinkurl{http://ascl.net/#1}}}
\providecommand{\doarXiv}[1]{\href{https://arxiv.org/abs/#1}{\nolinkurl{https://arxiv.org/abs/#1}}}

\bibitem[{{Adams} {et~al.}(2023){Adams}, {Conselice}, {Ferreira}, {Austin},
  {Trussler}, {Juod{\v{z}}balis}, {Wilkins}, {Caruana}, {Dayal}, {Verma}, \&
  {Vijayan}}]{Adams2023}
{Adams}, N.~J., {Conselice}, C.~J., {Ferreira}, L., {et~al.} 2023, \mnras, 518,
  4755, \dodoi{10.1093/mnras/stac3347}

\bibitem[{{Behroozi} {et~al.}(2020){Behroozi}, {Conroy}, {Wechsler}, {Hearin},
  {Williams}, {Moster}, {Yung}, {Somerville}, {Gottl{\"o}ber}, {Yepes}, \&
  {Endsley}}]{Behroozi2020}
{Behroozi}, P., {Conroy}, C., {Wechsler}, R.~H., {et~al.} 2020, \mnras, 499,
  5702, \dodoi{10.1093/mnras/staa3164}

\bibitem[{{Bouwens} {et~al.}(2022){Bouwens}, {Illingworth}, {Oesch},
  {Stefanon}, {Naidu}, {van Leeuwen}, \& {Magee}}]{Bouwens2022}
{Bouwens}, R., {Illingworth}, G., {Oesch}, P., {et~al.} 2022, arXiv e-prints,
  arXiv:2212.06683.
\newblock \doarXiv{2212.06683}

\bibitem[{{Boylan-Kolchin}(2022)}]{Boylan-Kolchin2022}
{Boylan-Kolchin}, M. 2022, arXiv e-prints, arXiv:2208.01611.
\newblock \doarXiv{2208.01611}

\bibitem[{{Crain} {et~al.}(2015){Crain}, {Schaye}, {Bower}, {Furlong},
  {Schaller}, {Theuns}, {Dalla Vecchia}, {Frenk}, {McCarthy}, {Helly},
  {Jenkins}, {Rosas-Guevara}, {White}, \& {Trayford}}]{Crain2015}
{Crain}, R.~A., {Schaye}, J., {Bower}, R.~G., {et~al.} 2015, \mnras, 450, 1937,
  \dodoi{10.1093/mnras/stv725}

\bibitem[{{Curtis-Lake} {et~al.}(2022){Curtis-Lake}, {Carniani}, {Cameron},
  {Charlot}, {Jakobsen}, {Maiolino}, {Bunker}, {Witstok}, {Smit}, {Chevallard},
  {Willott}, {Ferruit}, {Arribas}, {Bonaventura}, {Curti}, {D'Eugenio},
  {Franx}, {Giardino}, {Looser}, {L{\"u}tzgendorf}, {Maseda}, {Rawle}, {Rix},
  {Rodriguez del Pino}, {{\"U}bler}, {Sirianni}, {Dressler}, {Egami},
  {Eisenstein}, {Endsley}, {Hainline}, {Hausen}, {Johnson}, {Rieke},
  {Robertson}, {Shivaei}, {Stark}, {Tacchella}, {Williams}, {Willmer},
  {Bhatawdekar}, {Bowler}, {Boyett}, {Chen}, {de Graaff}, {Helton}, {Hviding},
  {Jones}, {Kumari}, {Lyu}, {Nelson}, {Perna}, {Sandles}, {Saxena}, {Suess},
  {Sun}, {Topping}, {Wallace}, \& {Whitler}}]{Curtis-Lake2022}
{Curtis-Lake}, E., {Carniani}, S., {Cameron}, A., {et~al.} 2022, arXiv
  e-prints, arXiv:2212.04568.
\newblock \doarXiv{2212.04568}

\bibitem[{{Dav{\'e}} {et~al.}(2019){Dav{\'e}}, {Angl{\'e}s-Alc{\'a}zar},
  {Narayanan}, {Li}, {Rafieferantsoa}, \& {Appleby}}]{Dave2019}
{Dav{\'e}}, R., {Angl{\'e}s-Alc{\'a}zar}, D., {Narayanan}, D., {et~al.} 2019,
  \mnras, 486, 2827, \dodoi{10.1093/mnras/stz937}

\bibitem[{{Dubois} {et~al.}(2014){Dubois}, {Pichon}, {Welker}, {Le Borgne},
  {Devriendt}, {Laigle}, {Codis}, {Pogosyan}, {Arnouts}, {Benabed}, {Bertin},
  {Blaizot}, {Bouchet}, {Cardoso}, {Colombi}, {de Lapparent}, {Desjacques},
  {Gavazzi}, {Kassin}, {Kimm}, {McCracken}, {Milliard}, {Peirani}, {Prunet},
  {Rouberol}, {Silk}, {Slyz}, {Sousbie}, {Teyssier}, {Tresse}, {Treyer},
  {Vibert}, \& {Volonteri}}]{Dubois2014}
{Dubois}, Y., {Pichon}, C., {Welker}, C., {et~al.} 2014, \mnras, 444, 1453,
  \dodoi{10.1093/mnras/stu1227}

\bibitem[{{Finkelstein} {et~al.}(2022){Finkelstein}, {Bagley}, {Ferguson},
  {Wilkins}, {Kartaltepe}, {Papovich}, {Yung}, {Arrabal Haro}, {Behroozi},
  {Dickinson}, {Kocevski}, {Koekemoer}, {Larson}, {Le Bail}, {Morales},
  {Perez-Gonzalez}, {Burgarella}, {Dave}, {Hirschmann}, {Somerville}, {Wuyts},
  {Bromm}, {Casey}, {Fontana}, {Fujimoto}, {Gardner}, {Giavalisco}, {Grazian},
  {Grogin}, {Hathi}, {Hutchison}, {Jha}, {Jogee}, {Kewley}, {Kirkpatrick},
  {Long}, {Lotz}, {Pentericci}, {Pierel}, {Pirzkal}, {Ravindranath}, {Ryan},
  {Trump}, {Yang}, {Bhatawdekar}, {Bisigello}, {Buat}, {Calabro}, {Castellano},
  {Cleri}, {Cooper}, {Croton}, {Daddi}, {Dekel}, {Elbaz}, {Franco}, {Gawiser},
  {Holwerda}, {Huertas-Company}, {Jaskot}, {Leung}, {Lucas}, {Mobasher},
  {Pandya}, {Tacchella}, {Weiner}, \& {Zavala}}]{Finkelstein2022}
{Finkelstein}, S.~L., {Bagley}, M.~B., {Ferguson}, H.~C., {et~al.} 2022, arXiv
  e-prints, arXiv:2211.05792.
\newblock \doarXiv{2211.05792}

\bibitem[{{Fujimoto} {et~al.}(2022){Fujimoto}, {Finkelstein}, {Burgarella},
  {Carilli}, {Buat}, {Casey}, {Ciesla}, {Tacchella}, {Zavala}, {Brammer},
  {Fudamoto}, {Ouchi}, {Valentino}, {Cooper}, {Dickinson}, {Franco},
  {Giavalisco}, {Hutchison}, {Kartaltepe}, {Koekemoer}, {Kojima}, {Larson},
  {Murphy}, {Papovich}, {P{\'e}rez-Gonz{\'a}lez}, {Somerville}, {Yoon},
  {Wilkins}, {Yung}, {Akins}, {Amor{\'\i}n}, {Arrabal Haro}, {Bagley},
  {Chworowsky}, {Cooper}, {Costantin}, {Daddi}, {Ferguson}, {Grogin},
  {Jim{\'e}nez-Andrade}, {Juneau}, {Kirkpatrick}, {Kocevski}, {Le Bail},
  {Long}, {Lucas}, {Magnelli}, {McKinney}, {Rose}, {Seill{\'e}}, {Simons}, \&
  {Weiner}}]{Fujimoto2022}
{Fujimoto}, S., {Finkelstein}, S.~L., {Burgarella}, D., {et~al.} 2022, arXiv
  e-prints, arXiv:2211.03896.
\newblock \doarXiv{2211.03896}

\bibitem[{{Genel} {et~al.}(2014){Genel}, {Vogelsberger}, {Springel}, {Sijacki},
  {Nelson}, {Snyder}, {Rodriguez-Gomez}, {Torrey}, \& {Hernquist}}]{Genel2014}
{Genel}, S., {Vogelsberger}, M., {Springel}, V., {et~al.} 2014, \mnras, 445,
  175, \dodoi{10.1093/mnras/stu1654}

\bibitem[{Harvey {et~al.}(1996)Harvey, Meeting, (1994), Brown, \&
  Smith}]{Harvey1996}
Harvey, P., Meeting, R. S. G. B.~D., (1994), R. S. G. B. D.~M., Brown, A., \&
  Smith, J. 1996, New Uses for New Phylogenies (Oxford University Press)

\bibitem[{{Haslbauer} {et~al.}(2022){Haslbauer}, {Kroupa}, {Zonoozi}, \&
  {Haghi}}]{Haslbauer2022}
{Haslbauer}, M., {Kroupa}, P., {Zonoozi}, A.~H., \& {Haghi}, H. 2022, \apjl,
  939, L31, \dodoi{10.3847/2041-8213/ac9a50}

\bibitem[{{Hinshaw} {et~al.}(2013){Hinshaw}, {Larson}, {Komatsu}, {Spergel},
  {Bennett}, {Dunkley}, {Nolta}, {Halpern}, {Hill}, {Odegard}, {Page}, {Smith},
  {Weiland}, {Gold}, {Jarosik}, {Kogut}, {Limon}, {Meyer}, {Tucker}, {Wollack},
  \& {Wright}}]{Hinshaw2013}
{Hinshaw}, G., {Larson}, D., {Komatsu}, E., {et~al.} 2013, \apjs, 208, 19,
  \dodoi{10.1088/0067-0049/208/2/19}

\bibitem[{{Johnson} {et~al.}(2021){Johnson}, {Leja}, {Conroy}, \&
  {Speagle}}]{Johnson2021}
{Johnson}, B.~D., {Leja}, J., {Conroy}, C., \& {Speagle}, J.~S. 2021, \apjs,
  254, 22, \dodoi{10.3847/1538-4365/abef67}

\bibitem[{{Kaasinen} {et~al.}(2022){Kaasinen}, {van Marrewijk}, {Popping},
  {Ginolfi}, {Di Mascolo}, {Mroczkowski}, {Concas}, {Di Cesare}, {Killi}, \&
  {Langan}}]{Kaasinen2022}
{Kaasinen}, M., {van Marrewijk}, J., {Popping}, G., {et~al.} 2022, arXiv
  e-prints, arXiv:2210.03754.
\newblock \doarXiv{2210.03754}

\bibitem[{{Komatsu} {et~al.}(2011){Komatsu}, {Smith}, {Dunkley}, {Bennett},
  {Gold}, {Hinshaw}, {Jarosik}, {Larson}, {Nolta}, {Page}, {Spergel},
  {Halpern}, {Hill}, {Kogut}, {Limon}, {Meyer}, {Odegard}, {Tucker}, {Weiland},
  {Wollack}, \& {Wright}}]{Komatsu2011}
{Komatsu}, E., {Smith}, K.~M., {Dunkley}, J., {et~al.} 2011, \apjs, 192, 18,
  \dodoi{10.1088/0067-0049/192/2/18}

\bibitem[{{Labbe} {et~al.}(2022){Labbe}, {van Dokkum}, {Nelson}, {Bezanson},
  {Suess}, {Leja}, {Brammer}, {Whitaker}, {Mathews}, \& {Stefanon}}]{Labbe2022}
{Labbe}, I., {van Dokkum}, P., {Nelson}, E., {et~al.} 2022, arXiv e-prints,
  arXiv:2207.12446.
\newblock \doarXiv{2207.12446}

\bibitem[{{Lovell} {et~al.}(2023){Lovell}, {Harrison}, {Harikane}, {Tacchella},
  \& {Wilkins}}]{Lovell2023}
{Lovell}, C.~C., {Harrison}, I., {Harikane}, Y., {Tacchella}, S., \& {Wilkins},
  S.~M. 2023, \mnras, 518, 2511, \dodoi{10.1093/mnras/stac3224}

\bibitem[{{Lovell} {et~al.}(2021){Lovell}, {Vijayan}, {Thomas}, {Wilkins},
  {Barnes}, {Irodotou}, \& {Roper}}]{Lovell2021}
{Lovell}, C.~C., {Vijayan}, A.~P., {Thomas}, P.~A., {et~al.} 2021, \mnras, 500,
  2127, \dodoi{10.1093/mnras/staa3360}

\bibitem[{{Marinacci} {et~al.}(2018){Marinacci}, {Vogelsberger}, {Pakmor},
  {Torrey}, {Springel}, {Hernquist}, {Nelson}, {Weinberger}, {Pillepich},
  {Naiman}, \& {Genel}}]{Marinacci2018}
{Marinacci}, F., {Vogelsberger}, M., {Pakmor}, R., {et~al.} 2018, \mnras, 480,
  5113, \dodoi{10.1093/mnras/sty2206}

\bibitem[{{McAlpine} {et~al.}(2016){McAlpine}, {Helly}, {Schaller}, {Trayford},
  {Qu}, {Furlong}, {Bower}, {Crain}, {Schaye}, {Theuns}, {Dalla Vecchia},
  {Frenk}, {McCarthy}, {Jenkins}, {Rosas-Guevara}, {White}, {Baes}, {Camps}, \&
  {Lemson}}]{McAlpine2016}
{McAlpine}, S., {Helly}, J.~C., {Schaller}, M., {et~al.} 2016, Astronomy and
  Computing, 15, 72, \dodoi{10.1016/j.ascom.2016.02.004}

\bibitem[{{Naab} \& {Ostriker}(2017)}]{Naab2017}
{Naab}, T., \& {Ostriker}, J.~P. 2017, \araa, 55, 59,
  \dodoi{10.1146/annurev-astro-081913-040019}

\bibitem[{{Naidu} {et~al.}(2022{\natexlab{a}}){Naidu}, {Oesch}, {Dokkum},
  {Nelson}, {Suess}, {Brammer}, {Whitaker}, {Illingworth}, {Bouwens},
  {Tacchella}, {Matthee}, {Allen}, {Bezanson}, {Conroy}, {Labbe}, {Leja},
  {Leonova}, {Magee}, {Price}, {Setton}, {Strait}, {Stefanon}, {Toft},
  {Weaver}, \& {Weibel}}]{Naidu2022a}
{Naidu}, R.~P., {Oesch}, P.~A., {Dokkum}, P.~v., {et~al.} 2022{\natexlab{a}},
  \apjl, 940, L14, \dodoi{10.3847/2041-8213/ac9b22}

\bibitem[{{Naidu} {et~al.}(2022{\natexlab{b}}){Naidu}, {Oesch}, {Setton},
  {Matthee}, {Conroy}, {Johnson}, {Weaver}, {Bouwens}, {Brammer}, {Dayal},
  {Illingworth}, {Barrufet}, {Belli}, {Bezanson}, {Bose}, {Heintz}, {Leja},
  {Leonova}, {Marques-Chaves}, {Stefanon}, {Toft}, {van der Wel}, {van Dokkum},
  {Weibel}, \& {Whitaker}}]{Naidu2022b}
{Naidu}, R.~P., {Oesch}, P.~A., {Setton}, D.~J., {et~al.} 2022{\natexlab{b}},
  arXiv e-prints, arXiv:2208.02794.
\newblock \doarXiv{2208.02794}

\bibitem[{{Naiman} {et~al.}(2018){Naiman}, {Pillepich}, {Springel},
  {Ramirez-Ruiz}, {Torrey}, {Vogelsberger}, {Pakmor}, {Nelson}, {Marinacci},
  {Hernquist}, {Weinberger}, \& {Genel}}]{Naiman2018}
{Naiman}, J.~P., {Pillepich}, A., {Springel}, V., {et~al.} 2018, \mnras, 477,
  1206, \dodoi{10.1093/mnras/sty618}

\bibitem[{{Nelson} {et~al.}(2015){Nelson}, {Pillepich}, {Genel},
  {Vogelsberger}, {Springel}, {Torrey}, {Rodriguez-Gomez}, {Sijacki}, {Snyder},
  {Griffen}, {Marinacci}, {Blecha}, {Sales}, {Xu}, \&
  {Hernquist}}]{Nelson2015b}
{Nelson}, D., {Pillepich}, A., {Genel}, S., {et~al.} 2015, Astronomy and
  Computing, 13, 12, \dodoi{10.1016/j.ascom.2015.09.003}

\bibitem[{{Nelson} {et~al.}(2018){Nelson}, {Pillepich}, {Springel},
  {Weinberger}, {Hernquist}, {Pakmor}, {Genel}, {Torrey}, {Vogelsberger},
  {Kauffmann}, {Marinacci}, \& {Naiman}}]{Nelson2018}
{Nelson}, D., {Pillepich}, A., {Springel}, V., {et~al.} 2018, \mnras, 475, 624,
  \dodoi{10.1093/mnras/stx3040}

\bibitem[{{Nelson} {et~al.}(2019{\natexlab{a}}){Nelson}, {Pillepich},
  {Springel}, {Pakmor}, {Weinberger}, {Genel}, {Torrey}, {Vogelsberger},
  {Marinacci}, \& {Hernquist}}]{Nelson2019b}
---. 2019{\natexlab{a}}, \mnras, 490, 3234, \dodoi{10.1093/mnras/stz2306}

\bibitem[{{Nelson} {et~al.}(2019{\natexlab{b}}){Nelson}, {Springel},
  {Pillepich}, {Rodriguez-Gomez}, {Torrey}, {Genel}, {Vogelsberger}, {Pakmor},
  {Marinacci}, {Weinberger}, {Kelley}, {Lovell}, {Diemer}, \&
  {Hernquist}}]{Nelson2019a}
{Nelson}, D., {Springel}, V., {Pillepich}, A., {et~al.} 2019{\natexlab{b}},
  Computational Astrophysics and Cosmology, 6, 2,
  \dodoi{10.1186/s40668-019-0028-x}

\bibitem[{{O'Shea} {et~al.}(2015){O'Shea}, {Wise}, {Xu}, \&
  {Norman}}]{O'Shea2015}
{O'Shea}, B.~W., {Wise}, J.~H., {Xu}, H., \& {Norman}, M.~L. 2015, \apjl, 807,
  L12, \dodoi{10.1088/2041-8205/807/1/L12}

\bibitem[{{Pillepich} {et~al.}(2018{\natexlab{a}}){Pillepich}, {Springel},
  {Nelson}, {Genel}, {Naiman}, {Pakmor}, {Hernquist}, {Torrey}, {Vogelsberger},
  {Weinberger}, \& {Marinacci}}]{Pillepich2018a}
{Pillepich}, A., {Springel}, V., {Nelson}, D., {et~al.} 2018{\natexlab{a}},
  \mnras, 473, 4077, \dodoi{10.1093/mnras/stx2656}

\bibitem[{{Pillepich} {et~al.}(2018{\natexlab{b}}){Pillepich}, {Nelson},
  {Hernquist}, {Springel}, {Pakmor}, {Torrey}, {Weinberger}, {Genel}, {Naiman},
  {Marinacci}, \& {Vogelsberger}}]{Pillepich2018b}
{Pillepich}, A., {Nelson}, D., {Hernquist}, L., {et~al.} 2018{\natexlab{b}},
  \mnras, 475, 648, \dodoi{10.1093/mnras/stx3112}

\bibitem[{{Planck Collaboration} {et~al.}(2014){Planck Collaboration}, {Ade},
  {Aghanim}, {Armitage-Caplan}, {Arnaud}, {Ashdown}, {Atrio-Barandela},
  {Aumont}, {Baccigalupi}, {Banday}, \& et~al.}]{Planck2014}
{Planck Collaboration}, {Ade}, P.~A.~R., {Aghanim}, N., {et~al.} 2014, \aap,
  571, A16, \dodoi{10.1051/0004-6361/201321591}

\bibitem[{{Planck Collaboration} {et~al.}(2016){Planck Collaboration}, {Ade},
  {Aghanim}, {Arnaud}, {Ashdown}, {Aumont}, {Baccigalupi}, {Banday},
  {Barreiro}, {Bartlett}, {Bartolo}, {Battaner}, {Battye}, {Benabed},
  {Beno{\^\i}t}, {Benoit-L{\'e}vy}, {Bernard}, {Bersanelli}, {Bielewicz},
  {Bock}, {Bonaldi}, {Bonavera}, {Bond}, {Borrill}, {Bouchet}, {Boulanger},
  {Bucher}, {Burigana}, {Butler}, {Calabrese}, {Cardoso}, {Catalano},
  {Challinor}, {Chamballu}, {Chary}, {Chiang}, {Chluba}, {Christensen},
  {Church}, {Clements}, {Colombi}, {Colombo}, {Combet}, {Coulais}, {Crill},
  {Curto}, {Cuttaia}, {Danese}, {Davies}, {Davis}, {de Bernardis}, {de Rosa},
  {de Zotti}, {Delabrouille}, {D{\'e}sert}, {Di Valentino}, {Dickinson},
  {Diego}, {Dolag}, {Dole}, {Donzelli}, {Dor{\'e}}, {Douspis}, {Ducout},
  {Dunkley}, {Dupac}, {Efstathiou}, {Elsner}, {En{\ss}lin}, {Eriksen},
  {Farhang}, {Fergusson}, {Finelli}, {Forni}, {Frailis}, {Fraisse},
  {Franceschi}, {Frejsel}, {Galeotta}, {Galli}, {Ganga}, {Gauthier}, {Gerbino},
  {Ghosh}, {Giard}, {Giraud-H{\'e}raud}, {Giusarma}, {Gjerl{\o}w},
  {Gonz{\'a}lez-Nuevo}, {G{\'o}rski}, {Gratton}, {Gregorio}, {Gruppuso},
  {Gudmundsson}, {Hamann}, {Hansen}, {Hanson}, {Harrison}, {Helou},
  {Henrot-Versill{\'e}}, {Hern{\'a}ndez-Monteagudo}, {Herranz}, {Hildebrandt},
  {Hivon}, {Hobson}, {Holmes}, {Hornstrup}, {Hovest}, {Huang}, {Huffenberger},
  {Hurier}, {Jaffe}, {Jaffe}, {Jones}, {Juvela}, {Keih{\"a}nen}, {Keskitalo},
  {Kisner}, {Kneissl}, {Knoche}, {Knox}, {Kunz}, {Kurki-Suonio}, {Lagache},
  {L{\"a}hteenm{\"a}ki}, {Lamarre}, {Lasenby}, {Lattanzi}, {Lawrence}, {Leahy},
  {Leonardi}, {Lesgourgues}, {Levrier}, {Lewis}, {Liguori}, {Lilje},
  {Linden-V{\o}rnle}, {L{\'o}pez-Caniego}, {Lubin}, {Mac{\'\i}as-P{\'e}rez},
  {Maggio}, {Maino}, {Mandolesi}, {Mangilli}, {Marchini}, {Maris}, {Martin},
  {Martinelli}, {Mart{\'\i}nez-Gonz{\'a}lez}, {Masi}, {Matarrese}, {McGehee},
  {Meinhold}, {Melchiorri}, {Melin}, {Mendes}, {Mennella}, {Migliaccio},
  {Millea}, {Mitra}, {Miville-Desch{\^e}nes}, {Moneti}, {Montier}, {Morgante},
  {Mortlock}, {Moss}, {Munshi}, {Murphy}, {Naselsky}, {Nati}, {Natoli},
  {Netterfield}, {N{\o}rgaard-Nielsen}, {Noviello}, {Novikov}, {Novikov},
  {Oxborrow}, {Paci}, {Pagano}, {Pajot}, {Paladini}, {Paoletti}, {Partridge},
  {Pasian}, {Patanchon}, {Pearson}, {Perdereau}, {Perotto}, {Perrotta},
  {Pettorino}, {Piacentini}, {Piat}, {Pierpaoli}, {Pietrobon}, {Plaszczynski},
  {Pointecouteau}, {Polenta}, {Popa}, {Pratt}, {Pr{\'e}zeau}, {Prunet},
  {Puget}, {Rachen}, {Reach}, {Rebolo}, {Reinecke}, {Remazeilles}, {Renault},
  {Renzi}, {Ristorcelli}, {Rocha}, {Rosset}, {Rossetti}, {Roudier},
  {Rouill{\'e} d'Orfeuil}, {Rowan-Robinson}, {Rubi{\~n}o-Mart{\'\i}n},
  {Rusholme}, {Said}, {Salvatelli}, {Salvati}, {Sandri}, {Santos},
  {Savelainen}, {Savini}, {Scott}, {Seiffert}, {Serra}, {Shellard}, {Spencer},
  {Spinelli}, {Stolyarov}, {Stompor}, {Sudiwala}, {Sunyaev}, {Sutton},
  {Suur-Uski}, {Sygnet}, {Tauber}, {Terenzi}, {Toffolatti}, {Tomasi},
  {Tristram}, {Trombetti}, {Tucci}, {Tuovinen}, {T{\"u}rler}, {Umana},
  {Valenziano}, {Valiviita}, {Van Tent}, {Vielva}, {Villa}, {Wade}, {Wandelt},
  {Wehus}, {White}, {White}, {Wilkinson}, {Yvon}, {Zacchei}, \&
  {Zonca}}]{Planck2016}
---. 2016, \aap, 594, A13, \dodoi{10.1051/0004-6361/201525830}

\bibitem[{{Robertson} {et~al.}(2022){Robertson}, {Tacchella}, {Johnson},
  {Hainline}, {Whitler}, {Eisenstein}, {Endsley}, {Rieke}, {Stark}, {Alberts},
  {Dressler}, {Egami}, {Hausen}, {Rieke}, {Shivaei}, {Williams}, {Willmer},
  {Arribas}, {Bonaventura}, {Bunker}, {Cameron}, {Carniani}, {Charlot},
  {Chevallard}, {Curti}, {Curtis-Lake}, {D'Eugenio}, {Jakobsen}, {Looser},
  {L{\"u}tzgendorf}, {Maiolino}, {Maseda}, {Rawle}, {Rix}, {Smit}, {{\"U}bler},
  {Willott}, {Witstok}, {Baum}, {Bhatawdekar}, {Boyett}, {Chen}, {de Graaff},
  {Florian}, {Helton}, {Hviding}, {Ji}, {Kumari}, {Lyu}, {Nelson}, {Sandles},
  {Saxena}, {Suess}, {Sun}, {Topping}, \& {Wallace}}]{Robertson2022}
{Robertson}, B.~E., {Tacchella}, S., {Johnson}, B.~D., {et~al.} 2022, arXiv
  e-prints, arXiv:2212.04480.
\newblock \doarXiv{2212.04480}

\bibitem[{{Rodighiero} {et~al.}(2011){Rodighiero}, {Daddi}, {Baronchelli},
  {Cimatti}, {Renzini}, {Aussel}, {Popesso}, {Lutz}, {Andreani}, {Berta},
  {Cava}, {Elbaz}, {Feltre}, {Fontana}, {F{\"o}rster Schreiber},
  {Franceschini}, {Genzel}, {Grazian}, {Gruppioni}, {Ilbert}, {Le Floch},
  {Magdis}, {Magliocchetti}, {Magnelli}, {Maiolino}, {McCracken}, {Nordon},
  {Poglitsch}, {Santini}, {Pozzi}, {Riguccini}, {Tacconi}, {Wuyts}, \&
  {Zamorani}}]{Rodighiero2011}
{Rodighiero}, G., {Daddi}, E., {Baronchelli}, I., {et~al.} 2011, \apjl, 739,
  L40, \dodoi{10.1088/2041-8205/739/2/L40}

\bibitem[{{Rosdahl} {et~al.}(2018){Rosdahl}, {Katz}, {Blaizot}, {Kimm},
  {Michel-Dansac}, {Garel}, {Haehnelt}, {Ocvirk}, \& {Teyssier}}]{Rosdahl2018}
{Rosdahl}, J., {Katz}, H., {Blaizot}, J., {et~al.} 2018, \mnras, 479, 994,
  \dodoi{10.1093/mnras/sty1655}

\bibitem[{{Santini} {et~al.}(2017){Santini}, {Fontana}, {Castellano}, {Di
  Criscienzo}, {Merlin}, {Amorin}, {Cullen}, {Daddi}, {Dickinson}, {Dunlop},
  {Grazian}, {Lamastra}, {McLure}, {Micha{\l}owski}, {Pentericci}, \&
  {Shu}}]{Santini2017}
{Santini}, P., {Fontana}, A., {Castellano}, M., {et~al.} 2017, \apj, 847, 76,
  \dodoi{10.3847/1538-4357/aa8874}

\bibitem[{{Schaye} {et~al.}(2015){Schaye}, {Crain}, {Bower}, {Furlong},
  {Schaller}, {Theuns}, {Dalla Vecchia}, {Frenk}, {McCarthy}, {Helly},
  {Jenkins}, {Rosas-Guevara}, {White}, {Baes}, {Booth}, {Camps}, {Navarro},
  {Qu}, {Rahmati}, {Sawala}, {Thomas}, \& {Trayford}}]{Schaye2015}
{Schaye}, J., {Crain}, R.~A., {Bower}, R.~G., {et~al.} 2015, \mnras, 446, 521,
  \dodoi{10.1093/mnras/stu2058}

\bibitem[{{Sheth} \& {Tormen}(1999)}]{Sheth1999}
{Sheth}, R.~K., \& {Tormen}, G. 1999, \mnras, 308, 119,
  \dodoi{10.1046/j.1365-8711.1999.02692.x}

\bibitem[{{Springel} {et~al.}(2018){Springel}, {Pakmor}, {Pillepich},
  {Weinberger}, {Nelson}, {Hernquist}, {Vogelsberger}, {Genel}, {Torrey},
  {Marinacci}, \& {Naiman}}]{Springel2018}
{Springel}, V., {Pakmor}, R., {Pillepich}, A., {et~al.} 2018, \mnras, 475, 676,
  \dodoi{10.1093/mnras/stx3304}

\bibitem[{{Stefanon} {et~al.}(2021){Stefanon}, {Bouwens}, {Labb{\'e}},
  {Illingworth}, {Gonzalez}, \& {Oesch}}]{Stefanon2021}
{Stefanon}, M., {Bouwens}, R.~J., {Labb{\'e}}, I., {et~al.} 2021, \apj, 922,
  29, \dodoi{10.3847/1538-4357/ac1bb6}

\bibitem[{{Trebitsch} {et~al.}(2021){Trebitsch}, {Dubois}, {Volonteri},
  {Pfister}, {Cadiou}, {Katz}, {Rosdahl}, {Kimm}, {Pichon}, {Beckmann},
  {Devriendt}, \& {Slyz}}]{Trebitsch2021}
{Trebitsch}, M., {Dubois}, Y., {Volonteri}, M., {et~al.} 2021, \aap, 653, A154,
  \dodoi{10.1051/0004-6361/202037698}

\bibitem[{{Tremmel} {et~al.}(2017){Tremmel}, {Karcher}, {Governato},
  {Volonteri}, {Quinn}, {Pontzen}, {Anderson}, \& {Bellovary}}]{Tremmel2017}
{Tremmel}, M., {Karcher}, M., {Governato}, F., {et~al.} 2017, \mnras, 470,
  1121, \dodoi{10.1093/mnras/stx1160}

\bibitem[{{Tremmel} {et~al.}(2019){Tremmel}, {Quinn}, {Ricarte}, {Babul},
  {Chadayammuri}, {Natarajan}, {Nagai}, {Pontzen}, \&
  {Volonteri}}]{Tremmel2019}
{Tremmel}, M., {Quinn}, T.~R., {Ricarte}, A., {et~al.} 2019, \mnras, 483, 3336,
  \dodoi{10.1093/mnras/sty3336}

\bibitem[{{Visbal} {et~al.}(2020){Visbal}, {Bryan}, \& {Haiman}}]{Visbal2020}
{Visbal}, E., {Bryan}, G.~L., \& {Haiman}, Z. 2020, \apj, 897, 95,
  \dodoi{10.3847/1538-4357/ab994e}

\bibitem[{{Vogelsberger} {et~al.}(2014{\natexlab{a}}){Vogelsberger}, {Genel},
  {Springel}, {Torrey}, {Sijacki}, {Xu}, {Snyder}, {Bird}, {Nelson}, \&
  {Hernquist}}]{Vogelsberger2014a}
{Vogelsberger}, M., {Genel}, S., {Springel}, V., {et~al.} 2014{\natexlab{a}},
  \nat, 509, 177, \dodoi{10.1038/nature13316}

\bibitem[{{Vogelsberger} {et~al.}(2014{\natexlab{b}}){Vogelsberger}, {Genel},
  {Springel}, {Torrey}, {Sijacki}, {Xu}, {Snyder}, {Nelson}, \&
  {Hernquist}}]{Vogelsberger2014b}
---. 2014{\natexlab{b}}, \mnras, 444, 1518, \dodoi{10.1093/mnras/stu1536}

\bibitem[{{Zavala} {et~al.}(2022){Zavala}, {Buat}, {Casey}, {Burgarella},
  {Finkelstein}, {Bagley}, {Ciesla}, {Daddi}, {Dickinson}, {Ferguson},
  {Franco}, {Jim'enez-Andrade}, {Kartaltepe}, {Koekemoer}, {Le Bail}, {Murphy},
  {Papovich}, {Tacchella}, {Wilkins}, {Aretxaga}, {Behroozi}, {Champagne},
  {Fontana}, {Giavalisco}, {Grazian}, {Grogin}, {Kewley}, {Kocevski},
  {Kirkpatrick}, {Lotz}, {Pentericci}, {Perez-Gonzalez}, {Pirzkal},
  {Ravindranath}, {Somerville}, {Trump}, {Yang}, {Yung}, {Almaini}, {Amorin},
  {Annunziatella}, {Arrabal Haro}, {Backhaus}, {Barro}, {Bell}, {Bhatawdekar},
  {Bisigello}, {Buitrago}, {Calabro}, {Castellano}, {Chavez Ortiz},
  {Chworowsky}, {Cleri}, {Cohen}, {Cole}, {Cooke}, {Cooper}, {Cooray},
  {Costantin}, {Cox}, {Croton}, {Dave}, {de la Vega}, {Dekel}, {Elbaz},
  {Estrada-Carpenter}, {Fern{\'a}ndez}, {Finkelstein}, {Freundlich},
  {Fujimoto}, {Garc{\'\i}a-Argum{\'a}nez}, {Gardner}, {Gawiser},
  {G{\'o}mez-Guijarro}, {Guo}, {Hamilton}, {Hathi}, {Holwerda}, {Hirschmann},
  {Huertas-Company}, {Hutchison}, {Iyer}, {Jaskot}, {Jha}, {Jogee}, {Juneau},
  {Jung}, {Kassin}, {Kurczynski}, {Larson}, {Leung}, {Long}, {Lucas},
  {Magnelli}, {Mantha}, {Matharu}, {McGrath}, {McIntosh}, {Medrano}, {Merlin},
  {Mobasher}, {Morales}, {Newman}, {Nicholls}, {Pandya}, {Rafelski}, {Ronayne},
  {Rose}, {Ryan}, {Santini}, {Seill{\'e}}, {Shah}, {Shen}, {Simons}, {Snyder},
  {Stanway}, {Straughn}, {Teplitz}, {Vanderhoof}, {Vega-Ferrero}, {Wang},
  {Weiner}, {Willmer}, \& {Wuyts}}]{Zavala2022}
{Zavala}, J.~A., {Buat}, V., {Casey}, C.~M., {et~al.} 2022, arXiv e-prints,
  arXiv:2208.01816.
\newblock \doarXiv{2208.01816}

\end{thebibliography}

\end{document}